\documentclass[conference]{IEEEtran}
\IEEEoverridecommandlockouts
\usepackage{cite}
\usepackage{amsmath,amssymb,amsfonts}
\usepackage{algorithmic}
\usepackage{graphicx}
\usepackage{textcomp}
\usepackage{xcolor}
\usepackage{caption}
\usepackage{subfigure}
\usepackage{acronym}
\usepackage[bookmarks,colorlinks]{hyperref}
\usepackage{algorithm,algorithmic}
\def\BibTeX{{\rm B\kern-.05em{\sc i\kern-.025em b}\kern-.08em
    T\kern-.1667em\lower.7ex\hbox{E}\kern-.125emX}}

\acrodef{rmse}[RMSE]{root-mean-square error}
\acrodef{cf}[CF]{constant frequency}
\acrodef{sf}[SF]{stepped frequency}
\acrodef{rsf}[RSF]{randomized stepped frequency}
\acrodef{hrrp}[HRRP]{high resolution range profile}
\acrodef{prf}[PRF]{pulse repetition frequency}
\acrodef{pri}[PRI]{pulse repetition interval}
\acrodef{eccm}[ECCM]{electromagnetic counter-countermeasures}
\acrodef{uav}[UAV]{unmanned aerial vehicle}
\acrodef{rcs}[RCS]{radar cross-section}
\acrodef{herm}[HERM]{helicopter rotation modulation}
\acrodef{snr}[SNR]{signal-to-noise ratio}
\acrodef{mse}[MSE]{Mean Squared Error}
\acrodef{mmse}[MMSE]{minimum mean square error}
\acrodef{atr}[ATR]{Automatic Target Recognition}
\acrodef{dnns}[DNN]{Deep Neural Network}
\acrodef{cnns}[CNN]{Convolutional Neural Network}
\acrodef{fan}[FAN]{Frequency Attention Network}
\acrodef{cfa}[CFA]{Complex-valued Frequency Attention}

\acrodef{rnns}[RNN]{Recurrent Neural Network}
\acrodef{drfm}[DRFM]{Digital Radio Frequency Memory}
\acrodef{lstm}[LSTM]{Long Short-Term Memory}
\acrodef{gru}[GRU]{gated recurrent unit}
\acrodef{psf}[PSF]{point spread function}
\acrodef{psd}[PSD]{Power Spectral Density}
\acrodef{fa}[FA]{Frequency Agile}
\acrodef{sjr}[SJR]{signal-to-jamming ratio}

\title{Robust Radar HRRP Recognition under Non-uniform Jamming Based on Complex-valued Frequency Attention Network 
\thanks{This work was supported by the National Natural Science Foundation of China under Grants 62301295 and the corresponding author is Lei Wang (e-mail: leiwangqh@tsinghua.edu.cn).}
}

\author{\IEEEauthorblockN{1\textsuperscript{st} Yanhao Wang , 2\textsuperscript{nd} Lei Wang*, and 3\textsuperscript{rd} Jie Wang and 4\textsuperscript{th} Yimin Liu}
\IEEEauthorblockA{\textit{Department of Electronic Engineering} \\
\textit{Tsinghua University}\\
Beijing, China \\
leiwangqh@tsinghua.edu.cn}

~\\
}
\vspace{-1.5cm}
\begin{document}

\maketitle

\begin{abstract}
Complex electromagnetic environments, often containing multiple jammers with different jamming patterns, produce non-uniform jamming power across the frequency spectrum. This spectral non-uniformity directly induces severe distortion in the target's \ac{hrrp}, consequently compromising the performance and reliability of conventional HRRP-based target recognition methods. This paper proposes a novel, end-to-end trained network for robust radar target recognition. The core of our model is a \ac{cfa} module that operates directly on the complex spectrum of the received echo. The \ac{cfa} module learns to generate an adaptive frequency-domain filter, assigning lower weights to bands corrupted by strong jamming while preserving critical target information in cleaner bands. The filtered spectrum is then fed into a classifier backbone for recognition. Experimental results on simulated \ac{hrrp} data with various jamming combinations demonstrate our method's superiority. Notably, under severe jamming conditions, our model achieves a recognition accuracy nearly 9\% higher than traditional model-based approaches, all while introducing negligible computational overhead. This highlights its exceptional performance and robustness in challenging jamming environments.
\end{abstract}


\maketitle

\section{Introduction}
\ac{atr} is a core function in modern radar systems with critical applications in both military and civilian domains. Among various techniques,  \ac{hrrp} has become a principal method for \ac{atr} due to its ability to capture fine-grained scattering information about a target's physical structure\cite{1453769,6553058}. While conventional recognition methods based on handcrafted features have been explored, they often struggle with high computational complexity and reliance on expert knowledge\cite{FENG2017379,1634818}. Consequently, deep learning models, with their powerful end-to-end feature extraction capabilities, have emerged as the dominant approach for \ac{hrrp} target recognition.

Prevailing deep learning methods, utilizing architectures such as \ac{cnns}\cite{9352965,zhang_cnn-assisted_2025}, \ac{rnns}\cite{rnn}, and Transformer\cite{jiang_ptrans_2024}, have demonstrated superior performance. However, their success is often predicated on the assumption of a jamming-free electromagnetic environment. The modern battlefield is far more complex, characterized by active jamming techniques designed to deceive or overwhelm radar systems.

Current radar anti-jamming approaches are often categorized into two main paths: active evasion at the system level and passive suppression at the signal processing level. Within system-level techniques, the \ac{fa} waveform is a conventional and effective strategy, and it is also widely used in target feature extraction and recognition tasks\cite{wang_rotorcraft_2024,zhao_enhanced_2025}. By pseudo-randomly hopping the carrier frequency between pulses, it can evade narrow-band tracking jammers or induce barrage jammers to split their power across a wide bandwidth, thereby reducing the jamming power density within any single pulse's bandwidth. However, when faced with high-power or wideband jamming sources, the efficacy of waveform agility alone may be insufficient to eliminate all jammings, resulting in significant residual jamming energy in the received echo.

This residual jamming presents a severe challenge for recognition algorithms, as it frequently exhibits a non-uniform power distribution across the frequency spectrum. This non-uniformity stems from multiple factors, including varying jammer strategies and the interaction between deceptive jamming techniques and the agile radar waveform itself. Such complex active jamming severely distorts the target's HRRP in two primary ways: first, by degrading the \ac{sjr}, which submerges or distorts critical scattering centers; and second, by causing the HRRPs of different targets to overlap in the feature space, thereby reducing discrimination performance\cite{yu_radar_2024}. Existing deep learning models, often developed assuming ideal conditions, demonstrate a significant decline in accuracy under these circumstances.

To address the recognition challenges posed by such residual, non-uniform jamming in FA radar systems, we propose a robust target recognition method based on a frequency attention network. The core of our method is to utilize the frequency-domain power of the jammed echo as auxiliary information to construct an attention mechanism. This mechanism adaptively assigns weights across the frequency domain, suppressing bands corrupted by high-power jamming while enhancing the target's information in cleaner spectral regions. This integrated approach is designed to significantly improve the robustness and recognition accuracy of HRRP-based ATR in complex electromagnetic jamming environments.

\section{PRELIMINARIES}
\label{prim}

To lay the groundwork for our proposed method, this section will first establish the fundamental signal models for \ac{fa} waveforms, covering both the target echo and the jamming signals. This mathematical formulation is essential for understanding the subsequent signal processing and the motivation of our recognition network.

\subsection{Signal Model}

We begin by defining the mathematical representation of the target's \ac{hrrp}. This subsection details the signal model of the radar echo received from a target under the \ac{fa} transmission scheme and outlines the procedure for generating the one-dimensional \ac{hrrp} image from this raw echo data.

To formulate the \ac{hrrp}, we begin by modeling the signal response with an ideal point target under the \ac{fa}  waveform scheme. This foundational model will serve as the basis for understanding the representation of extended targets, as they can be regarded as the linear sum of multiple point targets.

First, define the transmitted waveform for the $n$-th pulse, $x_n(t)$, as:
\begin{align}
x_{n}(t)=a_{n}(t)e^{j2\pi f_{n}t},
\end{align}
where $a_{n}(t)$ is the complex baseband waveform of the $n$-th pulse, $f_{n}$ is its carrier frequency, and $t$ denotes the fast-time dimension.

After down-conversion and low-pass filtering at the receiver, the resulting baseband signal, $\tilde{x}_{n}(t)$, can be expressed as:
\begin{align}
\tilde{x}_{n}(t)=a_{n}(t-\tau_{n})e^{-j2\pi f_{n}\tau_{n}}.
\end{align}

The round-trip delay of the echo from the $n$-th pulse is $\tau_{n}=\frac{2(R+nVT_{r})}{c}$. This delay is based on the "stop-and-go" model, where $R$ is the initial range to the target, $V$ is the target's radial velocity, $T_{r}$ is the \ac{pri}, and $c$ is the speed of light.

Next, we transform the received echo into the frequency domain using the Fourier Transform. 
\begin{align}
\tilde{X}_{n}(f) = A_{n}(f) e^{-j4\pi \frac{(f+f_{n})(R+nVT_{r})}{c}}.
\end{align}

To synthesize a high-resolution one-dimensional profile, two main steps are required: pulse compression and motion compensation, for which the specific algorithms are detailed in \cite{6212202}. After these operations, the compensated frequency spectrum of the $n$-th pulse, $Y_n'(f)$, can be expressed as:
\begin{align}
Y_{n}'(f) = Y_{n}(f) \cdot e^{j4\pi \frac{(f+f_{n})nVT_{r}}{c}} = W_{n}(f)e^{-j4\pi \frac{(f+f_{n})R}{c}},
\end{align}
where $W_{n}(f)$ is the energy spectral density of the baseband pulse.

Finally, to form the full synthetic wideband spectrum, $\tilde{Y}(f)$, the compensated spectral data $Y_n'(f)$ from all pulses are coherently combined. This process arranges the spectrum of each pulse according to its carrier frequency $f_n$ to form a continuous, high-bandwidth spectrum.
The target's \ac{hrrp} is then obtained by taking the Inverse Fourier Transform ($\mathcal{F}^{-1}$) of this synthesized wideband spectrum:
\begin{align}
Y_{\text{HRRP}}(t) = \mathcal{F}^{-1}\{\tilde{Y}(f)\}.
\end{align}

This resulting $Y_{\text{HRRP}}(t)$ provides a one-dimensional scattering profile of the target along the radar's line-of-sight.

\subsection{Compound Jamming Model}
In a complex electronic warfare environment, a radar system is often confronted not by a single jamming source, but by a coordinated or uncoordinated collective of multiple jammers. To analyze radar performance under such conditions, we define a comprehensive mathematical model for the jamming signal, $J_R(t)$, at the radar's receiving aperture. This model accounts for the presence of $N$ distinct narrow-band noise jammers, each characterized by different parameters.

The total received jamming signal, $J_R(t)$, is modeled as the linear superposition of signals from $N$ independent jammers. Assuming each jammer emits a zero-mean, stationary Gaussian noise signal, the composite signal is given by:
\begin{align}
    J_R(t) = \sum_{i=1}^{N} J_{R,i}(t),
\end{align}
where $J_{R,i}(t)$ represents the jamming contribution from the $i$-th jammer. The power of the signal received from each individual jammer is fundamentally dependent on the propagation path, most critically its distance from the radar. The variance, $\sigma_{j,i}^2$, which represents the average power of the $i$-th jamming signal at the receiver, is determined by the one-way radar range equation:
\begin{align}
\sigma_{j,i}^2 = \frac{P_{JT,i} G_{J,i} G_{R,i} \lambda^2}{(4\pi R_{J,i})^2}.
\end{align}
Here, $P_{JT,i}$, $G_{J,i}$, and $R_{J,i}$ are the transmit power, antenna gain, and range associated with the $i$-th jammer, respectively, while $G_{R,i}$ is the radar's antenna gain in the direction of that jammer.

\begin{figure}
\centering
{\includegraphics[width=14pc]{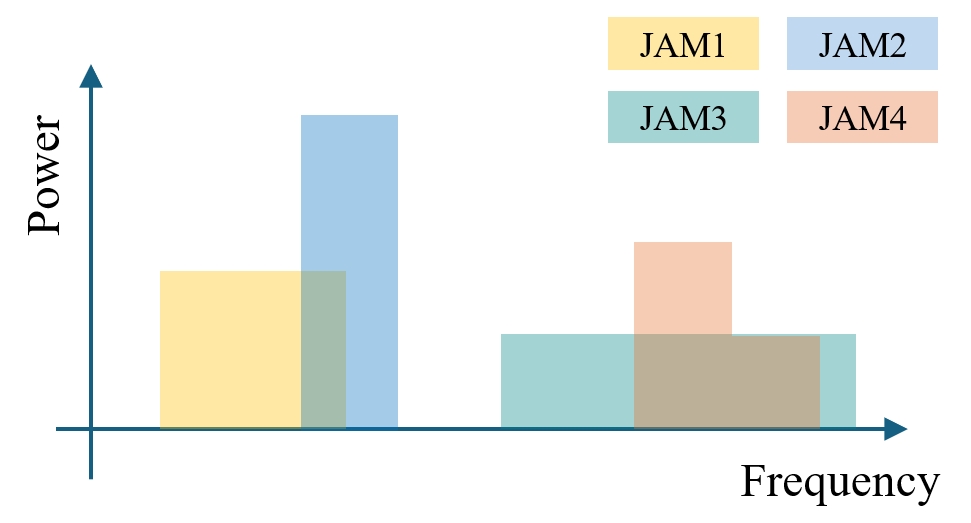}}
\caption{In the multi-jamming scenario, the \ac{psd} of the jamming signals received by the radar.}
\label{jam_diagram}
\end{figure}

It's worth noting that, since the individual jamming signals are statistically independent, the \ac{psd} of the composite signal $J_R(t)$ is the summation of the \ac{psd}s of each jamming signal, as shown in Fig.~\ref{jam_diagram}. The \ac{psd} for the $i$-th jammer, $S_{J,i}(f)$, is modeled as a rectangular function centered at its specific jamming frequency $f_{j,i}$ with a bandwidth of $B_{j,i}$. Consequently, the total \ac{psd}, $S_J(f)$, is expressed as:
\begin{align}
S_J(f) = \sum_{i=1}^{N} S_{J,i}(f),
\end{align}

where $ S_{J,i}(f) = \begin{cases} \frac{\sigma_{j,i}^2}{B_{j,i}} & \text{if } |f - f_{j,i}| \le \frac{B_{j,i}}{2} \\ 0 & \text{otherwise} \end{cases}$.

This formulation reveals a critical characteristic of the multi-jammer scenario. Because each jammer has a distinct power level ($\sigma_{j,i}^2$, determined by its range $R_{J,i}$ and other parameters), center frequency ($f_{j,i}$), and bandwidth ($B_{j,i}$), which results in a form of compound jamming. The total \ac{psd}, i.e., $S_J(f)$, is not flat across the radar's operational band. Instead, it exhibits a non-uniform profile, composed of multiple rectangular steps of varying heights and widths. The non-uniform spectrum under multi-jammer conditions poses significant challenges to conventional noise suppression techniques, necessitating more advanced adaptive processing and anti-jamming strategies to characterize and mitigate such complex jamming.

\section{Methodology}
\label{methods}

To motivate our deep learning solution, we first analyze the problem from a model-based perspective. By formally modeling the generation of a target's \ac{hrrp} and the distortion of non-uniform jamming, we establish the physical and mathematical basis for our proposed \ac{cfa}. The core principle of the method is that the frequency-domain power spectrum of the received echo contains valuable information about the energy distribution of both the target and the jamming. Our network leverages this as prior information, employing an attention mechanism to automatically suppress frequency bands occupied by strong jamming while enhancing important target features. This mechanism improves feature discrimination and enable robust, high-accuracy target recognition under strong jamming conditions.

 \subsection{Model-based Method}
We take the Wiener filter as an example of a model-based approach\cite{1643650}. It ensures the shape and structure similarity between the estimated signal and the ground truth since its optimality under the \ac{mmse} criterion.

Given an observed HRRP $x(n)$ composed of the clean signal $s(n)$ and additive non-uniform jamming $j(n)$, such that $x(n) = s(n) + j(n)$, the goal is to find an estimate $\hat{s}(n)$ that minimizes the \ac{mse}, defined as $E \left[|s(n) - \hat{s}(n)|^2\right]$.
The optimal linear filter that achieves this minimization operates in the frequency domain. The estimated signal spectrum $\hat{S}(k)$ is obtained by applying the Wiener filter's transfer function, $H(k)$, to the observed spectrum $X(k)$. The expression for this optimal filter is:
\begin{align}
	H(k) = \frac{P_s(k)}{P_s(k) + P_j(k)},
\end{align}
where $P_s(k)$ and $P_j(k)$ are the \ac{psd}s of the signal and jamming, respectively.
This equation elegantly demonstrates the filter's optimality under the \ac{mmse} criterion. The filter gain at each frequency bin $k$ is precisely determined by the local \ac{snr}, as captured by the PSDs. It optimally trades off jamming suppression with signal preservation—passing frequencies where the signal is strong and attenuating those dominated by jamming. The theoretical optimality of the Wiener filter is thus guaranteed, provided that these second-order statistics are accurately known.

Model-based methods suffer from two inherent limitations in complex jamming scenarios. These  filters depend on precise a priori statistical knowledge of the signal and jamming, information that is often unavailable in dynamic electromagnetic environments. Furthermore, they suffer from an inconsistency between signal-level and task-level goals, as they optimize for signal fidelity metrics (e.g., \ac{mse}) rather than the feature separability crucial for classification. This focus on signal-level optimality can inadvertently suppress weak yet highly discriminative features, ultimately undermining the performance of the downstream classification task.

\subsection{The Structure of the Network}
To realize the adaptive filtering strategy mentioned above, we designed the \ac{fan}, the architecture of which is detailed in this section. The cornerstone of \ac{fan} is the \ac{cfa}, a mechanism that functions as a differentiable, data-driven adaptive frequency filter.

\begin{figure}
\centering
{\includegraphics[width=90mm]{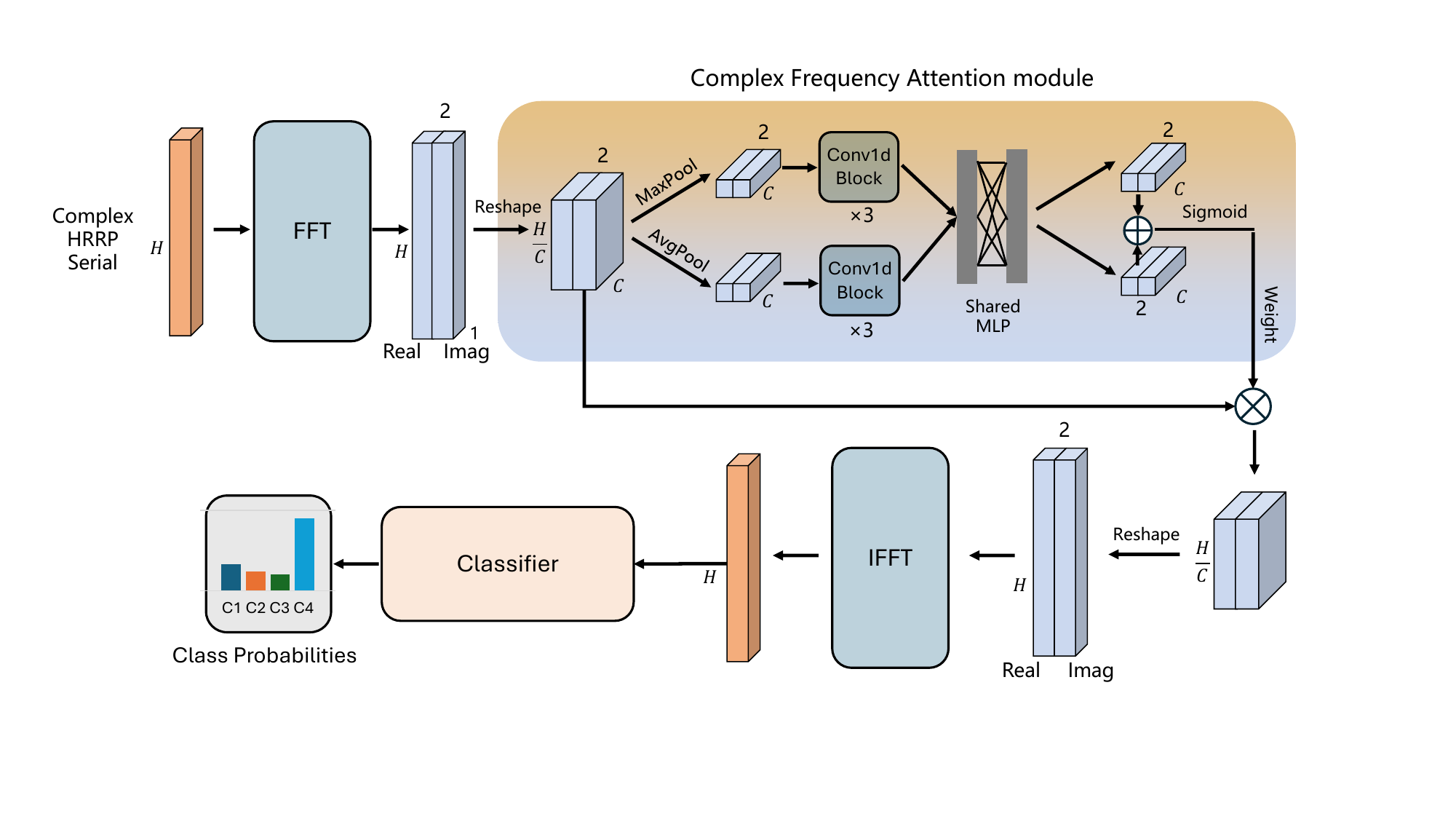}}
\caption{The network overall architecture and structure of key building blocks.}
\label{network-arc}
\end{figure}

The overall network architecture consists of two main stages, as shown in Fig.~\ref{network-arc}.

\begin{enumerate}
\item Attention-based Adaptive Filtering: The complex spectrum of HRRP is fed into the \ac{cfa} module, which outputs a weight vector. The core design of the CFA module first reshapes the 1D frequency spectrum into a multi-channel format. It then summarizes each channel's information using both Max and Average Pooling in parallel, creating a richer feature description by capturing both peak values and overall energy. These summary features are processed by 1D convolution blocks to model relationships between adjacent channels, followed by a shared MLP that computes the final attention weights. This process allows the module to adaptively amplify relevant frequency bands and suppress irrelevant ones based on learned inter-channel dependencies. The learned attention weights are then multiplied element-wise with the complex spectrum to perform adaptive filtering.
\item Classification: The weighted complex spectrum is then transformed into time domain and processed by a subsequent classifier backbone (e.g., a ResNet 1D\cite{ASNair20221DResNetFF} or Transformer\cite{10068427}) to produce the final class probabilities.
\end{enumerate}

As a data-driven method, the proposed frequency attention mechanism requires no explicit statistical models. The attention module implicitly learns the jamming's characteristics and generates a matching suppression strategy through end-to-end training on large datasets. More importantly, the filtering behavior is governed by signal fidelity and separability for classification. This task-driven optimization ensures that the learned feature representation is optimal at a semantic level, not just a signal level, thereby improving the robustness and accuracy of target recognition in complex jamming environments.

\section{Experiments}

\subsection{HRRP Dataset and Experimental Setup}
\begin{figure}
	\centering
	{\includegraphics[width=80mm]{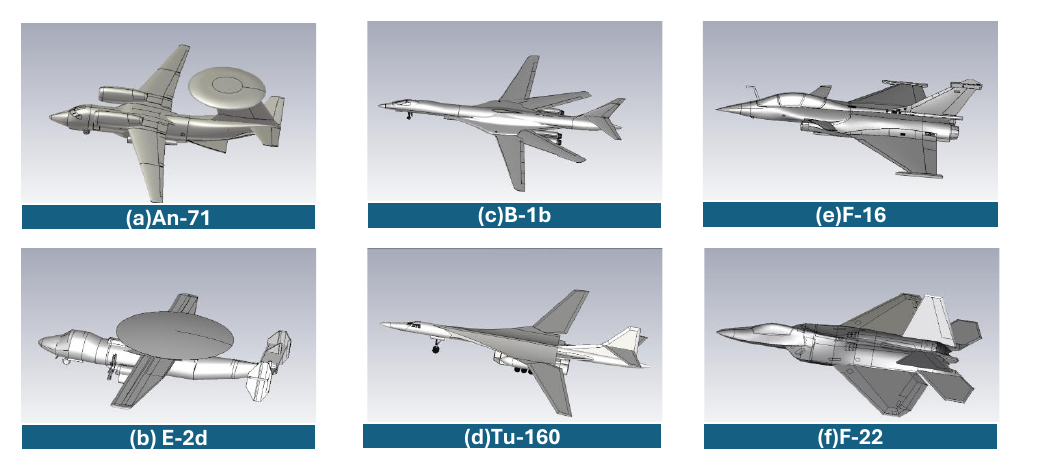}}
	\caption{Six classes of typical aircraft for generating multi-angle HRRP through CST Studio Suite.}
	\label{dataset}
\end{figure}

To validate the effectiveness of the proposed method, we constructed a simulated S-band HRRP dataset comprising six classes of typical aircraft: the An-71, E-2d, Tu-160, B-1b, F-16, and F-22, as shown in Fig.~\ref{dataset}. The generation process began by computing the \ac{rcs} of each aircraft model using the commercial electromagnetic simulation software, CST Studio Suite, across elevation angles from 0° to 5° and the full 360° of azimuth. Subsequently, based on the signal model detailed in Section~\ref{prim}, the RCS data was used to synthesize the HRRP samples. The synthetic bandwidth for all HRRPs was set to 800 MHz, spanning the frequency range from 2.4 GHz to 3.2 GHz, which results in a range resolution of 0.1875 m. A total of 1,800 samples were generated for each aircraft class. In the experimental setup for the \ac{fa} waveform, the number of available carrier frequencies was set to 16, with a corresponding sub-band bandwidth of 50 MHz. The compound jamming was added to the clean raw echoes across 8 of the 16 frequency bands.

For the experiments, All samples were randomly partitioned into non-overlapping training and testing sets with an 8:2 split. Our model was trained in an end-to-end manner for a total of 120 epochs, with a batch size of 64. We employed the Adam optimizer for training. All experiments were conducted on a single NVIDIA GeForce RTX 3090 GPU with 24 GB of memory.

\subsection{Experiment Result and Discussion}
In this section, we present a comprehensive evaluation of the proposed \ac{cfa} module, we conducted a series of comparative experiments. We integrated the \ac{cfa} module with two distinct deep learning backbones, a 1D ResNet and a Transformer, and benchmarked their performance against two sets of baseline models. The first baseline replaces our \ac{cfa} module with a traditional model-based Wiener filter to represent a classic anti-jamming approach. The second baseline, used for ablation studies, involves the standalone ResNet and Transformer classifiers operating directly on the jammed HRRP data, establishing a performance lower bound. All models were evaluated based on their average recognition accuracy (Avg. Acc), while their efficiency was compared using the number of learnable parameters (Params) and computational complexity (FLOPs).

\begin{figure}
	
	\centering
	{\includegraphics[width=80mm]{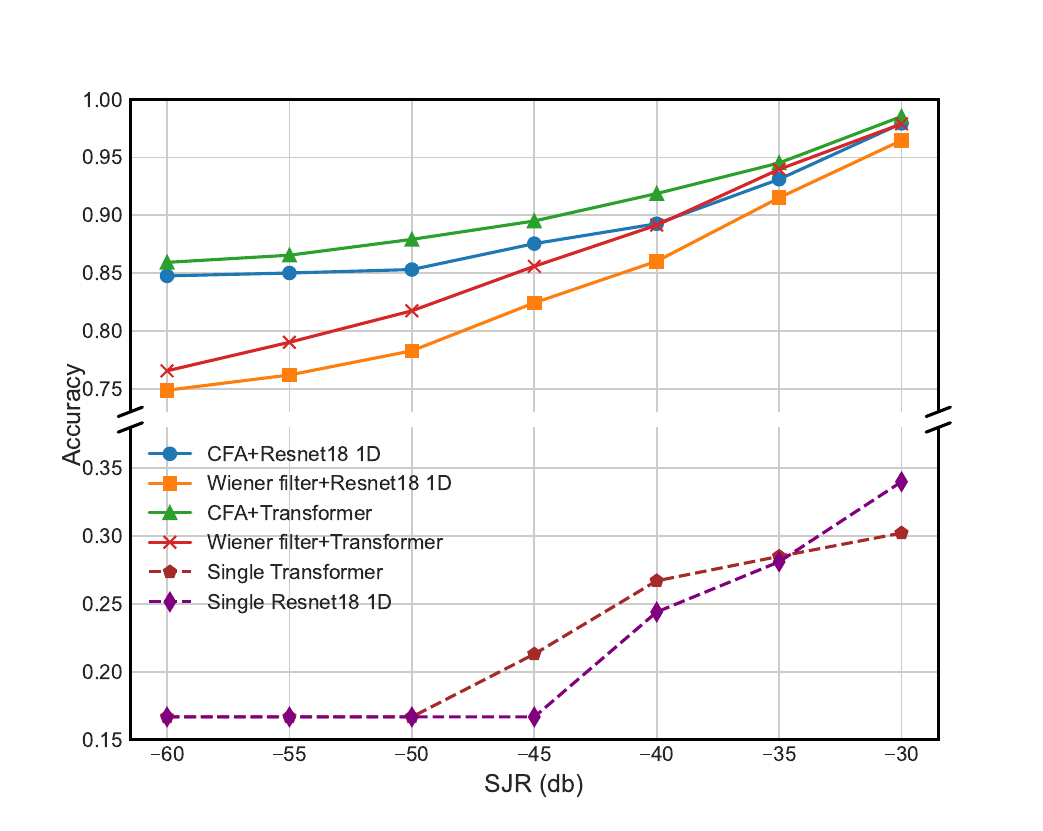}}
	\caption{Model accuracy under different \ac{sjr}.}
	\label{lines}
\end{figure}

To first establish the necessity of a dedicated anti-jamming module, we evaluated the performance of standalone classifiers on jammed HRRP data, as shown in Fig.~\ref{lines}. The results showed that the performance of both the ResNet 1D and Transformer models degrades significantly as the \ac{sjr} decreases. At an \ac{sjr} of -50dB and below, their accuracy plummets to approximately 16.7\%, which is equivalent to random guessing for a six-class problem. This outcome unequivocally demonstrates the vulnerability of modern deep learning classifiers to strong jamming, confirming that a front-end signal enhancement module is essential.

We then compared our proposed \ac{cfa} module against the traditional Wiener filter across a wide range of \ac{sjr}s, from -30dB to -60dB. The results clearly indicate that models equipped with our \ac{cfa} module consistently outperform their Wiener filter counterparts at all \ac{sjr} levels. Notably, this performance gap widens as the jamming becomes more severe. For instance, with the ResNet 1D backbone, \ac{cfa}'s accuracy advantage over the Wiener filter grows from 1.47\% at -30dB \ac{sjr} to a substantial 9.89\% in the low \ac{sjr} environment of -60dB. This trend confirms the superior robustness of our data-driven attention mechanism in challenging jamming conditions.

\begin{figure}
	\centering
	\subfigure[CFA+ResNet 1D]{
		\label{CM_CFA}
		\includegraphics[width=9pc]{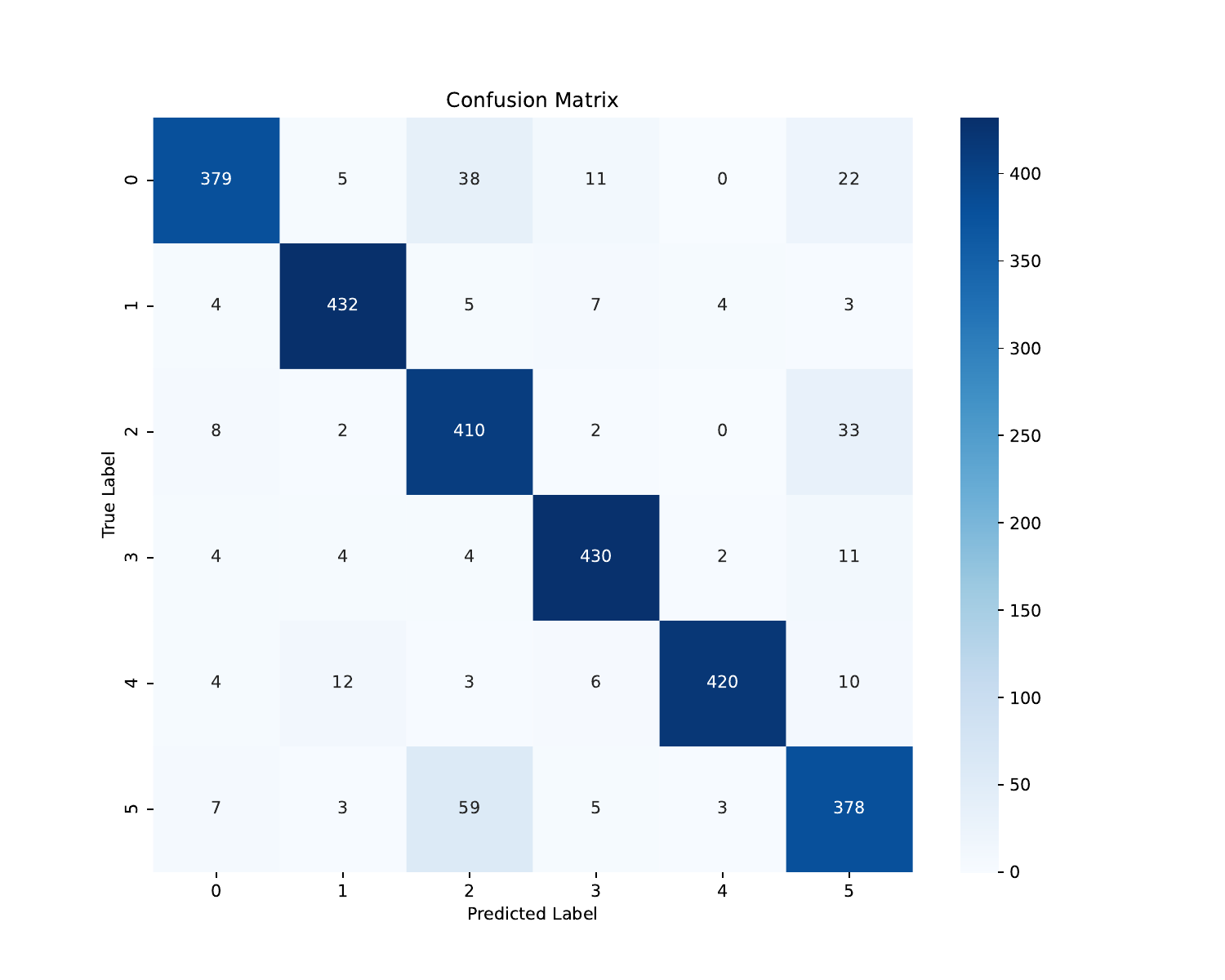}}
	\subfigure[Wiener filter+ResNet 1D]{
		\label{CM_model}
		\includegraphics[width=9pc]{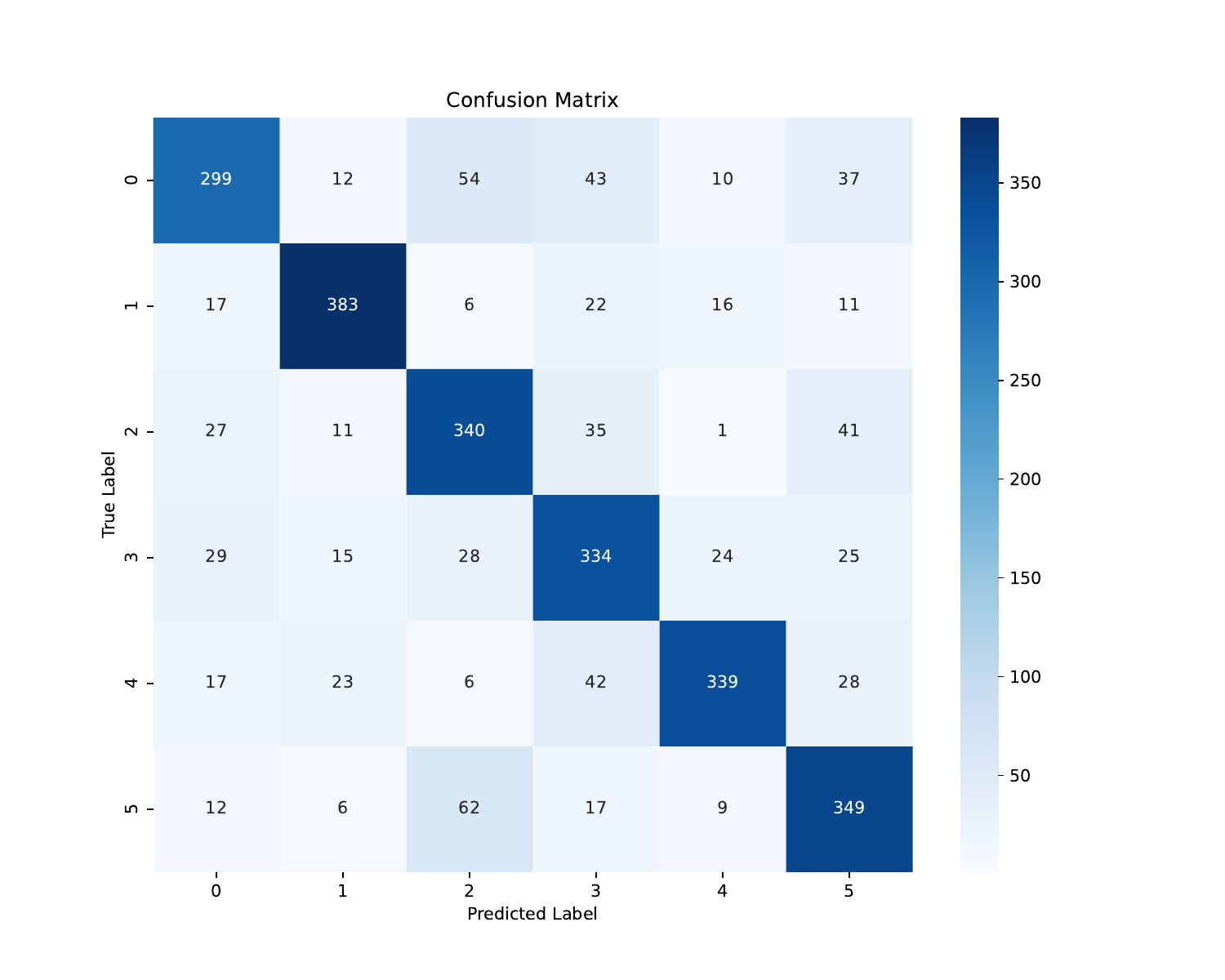}
	}
	\caption{The confusion matrices of the network with \ac{cfa} and Wiener filter.}
	\label{FIG_CM}
\end{figure}

An analysis and comparision among different models at -50dB \ac{sjr} further highlights the superiority and efficiency of the \ac{cfa} module. In Table~\ref{tab1}, the \ac{cfa}+ResNet 1D model achieved 85.31\% accuracy, a 7.03\% points higher than the Wiener filter solution. Similarly, the \ac{cfa}+Transformer model outperformed its Wiener counterpart by 6.18\% points. Critically, this substantial performance gain is achieved with a negligible increase in computational resources, as the \ac{cfa} module adds only 0.01 million parameters and less than 1\% to the total FLOPs. The confusion matrices in Fig.~\ref{FIG_CM} offer deeper insight, showing that the \ac{cfa}-based model maintains high accuracy across all classes with minimal confusion. In contrast, the Wiener filter-based model suffers from significant inter-class confusion, confirming that its extracted features are less discriminative. This comparison demonstrates that our \ac{cfa} module more effectively preserves target features while suppressing jamming, leading to a more robust feature representation and, consequently, higher recognition accuracy.

\begin{table}[h]
	
	\centering
	\caption{Comparision of the models at \ac{sjr}=-50dB} 
	\label{tab:model_comparison_simple_no_desc}
	\begin{tabular}{|l|c|c|c|}
		\hline \textbf{Model}
		& \textbf{Avg.Acc} & \textbf{Params ($10^6$)} & \textbf{FLOPs ($10^6$)} \\
		\hline
		\textbf{CFA}+ResNet 1D        & 85.31\%            & 2.19                     & 126.8                  \\
		\hline
		Wiener filter+ResNet 1D        & 78.28\%            & 2.18                    & 126.0                \\
		\hline
		\textbf{CFA}+Transformer       & 87.92\%            & 5.54                     & 2201.2                  \\
		\hline
		Wiener filter+Transformer & 81.74\%            & 5.53                     & 2151.5                 \\
		\hline
	\end{tabular}
	\label{tab1}
\end{table}

\section{Conclusion}
In this work, we proposed a network with \ac{cfa} module that achieves superior HRRP recognition performance under severe non-uniform jamming, significantly outperforming traditional methods with negligible computational cost. Looking forward, the lightweight and modular nature of the \ac{cfa} module makes it highly suitable for fine-tuning pre-trained classifiers, enabling rapid adaptation to dynamic jamming environments. Second, the attention weights learned by the CFA module and its filtered output offer significant interpretability into the model's anti-jamming mechanism. We plan to leverage visualization techniques in future work to analyze and showcase this transparency, further validating the network's decision-making process.

\label{conclu}

\bibliographystyle{unsrt}
\bibliography{ref.bib}

\end{document}